\documentclass[twocolumn,aps,prb]{revtex4}
\usepackage{graphicx}
\usepackage{amssymb}
\usepackage{amsmath}
\usepackage{bm}

\def\Journal #1,#2,#3,#4#5#6#7{#1 {\bf #2}, #3 (#4#5#6#7)}

\def\e{\varepsilon}

\begin{document}

\title{Orbital magnetism of graphene flakes}
\author{Yuya Ominato and Mikito Koshino}
\affiliation{Department of Physics, Tohoku University, Sendai 980-8578, Japan}
\date{\today}

\begin{abstract}
Orbital magnetism is studied for
graphene flakes with various shapes and edge configurations
using the tight-binding approximation.
In the low-temperature regime where 
the thermal energy is much smaller than
to the energy level spacing,
the susceptibility rapidly changes between diamagnetism
and paramagnetism as a function of Fermi energy,
in accordance with the energy level structure.
The susceptibility at charge neutral point is generally 
larger in armchair flake than in zigzag flake,
and larger in hexagonal flake than in triangular flake.
As the temperature increases, the discrete structures 
due to the quantum confinement are all gone,
and the susceptibility approximates the bulk limit
independently of the atomic configuration.
The diamagnetic current circulates entirely
on the graphene flake at zero temperature, 
while in increasing temperature it is localized near the edge 
with the characteristic depth proportional to $1/T$.
We predict that the diamagnetism of graphene can be observed 
using the alignment of graphene flakes 
in a feasible range of magnetic field.
\end{abstract}

\maketitle

\section{Introduction}

\begin{figure}
\begin{center}
\leavevmode\includegraphics[width=1.0\hsize]{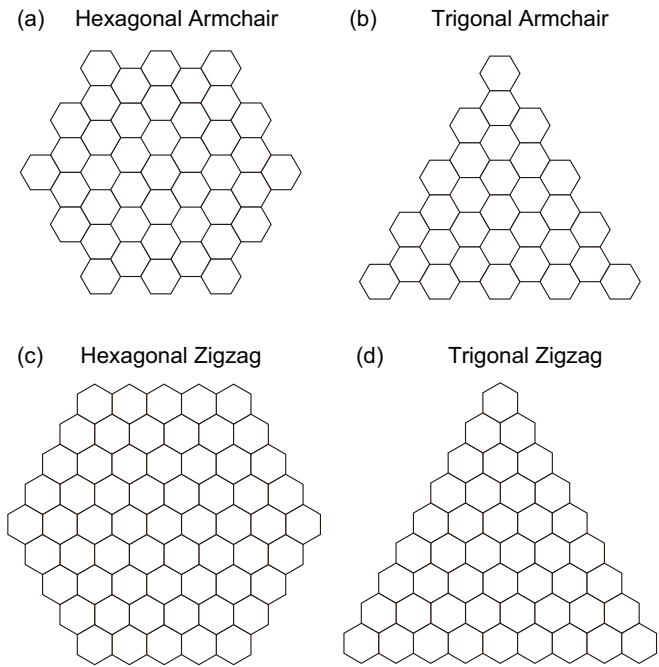}
\end{center}
\caption{Atomic structures of (a) hexagonal armchair,
(b) trigonal armchair, (c) hexagonal zigzag, and (d) trigonal zigzag 
graphene flakes.
}
\label{fig_flake}
\end{figure}

The recent developments in fabrication of graphene-based systems
realized a variety of graphene nanostructures,
such as graphene ribbons 
\cite{Han_et_al_2007a,Chen_et_al_2007a,Li_et_al_2008a,Kosynkin_et_al_2009a,Jiao_et_al_2009a}
and graphene flakes.
\cite{Kobayashi_et_al_2005a,Enoki_et_al_2007a,Geng_et_al_2012a,Hamalainen_et_al_2011a,Subramaniam_et_al_2012a}
The electronic band structure in these systems crucially 
depends on the shape and also on the edge termination,
\cite{Kobayashi_et_al_2005a,Fujita_et_al_1996a,Nakada_et_al_1996a,Wakabayashi_et_al_1999a}
giving physical properties distinct from those in bulk graphene.
So far, a number of theoretical researches have been devoted to 
understanding the electronic properties of graphene ribbons 
\cite{Fujita_et_al_1996a,Nakada_et_al_1996a,Wakabayashi_et_al_1999a,Ezawa_2006a,Brey_and_Fertig_2006a,Brey_and_Fertig_2006b,Son_et_al_2006a,Son_et_al_2006b}
and graphene flakes 
\cite{Ezawa_2007a,Ezawa_2010a,Zarenia_et_al_2011a,Potasz_et_al_2010a,Bahamon_et_al_2009a,
Akola_et_al_2008a,Zhang_and_Chang_2008a,Kosimov_et_al_2010a}
with various atomic configurations.

The purpose of this paper is to study the orbital magnetism of 
graphene flakes. 
Experimentally, the magnetic property
of graphene-based materials was investigated
for bulk graphite 
\cite{Krishnan_et_al_1937a,Esquinazi_et_al_2002a,Esquinazi_et_al_2003a},
nanographite \cite{Enoki_and_Takai_2009a},
and exfoliated graphene nanocrystals \cite{Sepioni_et_al_2010a}.
There the susceptibility always contains a strong diamagnetic background
due to the orbital effect,
whereas it is also contributed by the spin paramagnetism,
\cite{Sepioni_et_al_2010a}
and in some cases the spontaneous spin magnetic ordering
\cite{Esquinazi_et_al_2002a,Esquinazi_et_al_2003a,Enoki_and_Takai_2009a,Wang_et_al_2009a}
which can be caused by the zero-energy edge states
\cite{Wakabayashi_et_al_1999a,Fernandez_and_Palacios_2007a,Wang_and_Meng_and_Kaxiras_2008a}
and atomic defects.
In any case, correct understanding of the orbital susceptibility of 
finite graphene systems is important 
to describe the overall magnetic property in realistic graphene systems.

Graphene has unusual electronic band structure
characterized by the massless Dirac spectrum,
\cite{Novoselov_et_al_2005a,Zhang_et_al_2005a,Wallace_1947a,McClure_1956a,Slonczewski_and_Weiss_1958a,
DiVincenzo_and_Mele_1984a,Semenoff_1984a,Ando_2005a,Shon_and_Ando_1998a}
and accordingly, the orbital magnetism 
is significantly different from the conventional Landau diamagnetism.
\cite{McClure_1956a,McClure_1960a,Safran_and_DiSalvo_1979a,Blinowski_and_Rigaux_1984a,Saito_and_Kamimura_1986a,
Sharapov_et_al_2004a,Fukuyama_2007a,Nakamura_2007a,
Koshino_and_Ando_2007b,Ghosal_et_al_2007a,
Ando_2007d,Koshino_et_al_2009a,Koshino_and_Ando_2010a}
The orbital susceptibility of bulk graphene
diverges when the Fermi energy resides at Dirac point, but vanishes inside 
the conduction or the valence band.
Finite-size effect on this singular diamagnetism
has been theoretically studied for carbon nanotubes 
\cite{Ajiki_and_Ando_1993b,Ajiki_and_Ando_1995c,Yamamoto_et_al_2008a} 
and graphene ribbons.
\cite{Wakabayashi_et_al_1999a,Ominato_and_Koshino_2012a}
In our previous work, \cite{Ominato_and_Koshino_2012a} particularly,
we found that the susceptibility of graphene ribbon behaves 
in a complicated manner as a function of Fermi energy,
reflecting the subband quantization imposed by the spacial confinement.

In this paper, we consider the orbital diamagnetism of 
lower dimensional systems --- graphene flakes
as illustrated in Fig.\ \ref{fig_flake}.
For each case we calculate the orbital magnetic susceptibility
and the diamagnetic electric current distribution using the tight-binding model.
We find characteristic properties peculiar to each different case,
and also general tendencies independent of the configuration.
We also predict that the diamagnetism of graphene can be observed 
using the alignment of graphene flakes dissolved in a solvent
under a magnetic field.
The paper is organized as follows. In Sec.\ \ref{sec_form},
we introduce tight-binding Hamiltonian and the formulas
to describe the orbital magnetic effect.
We calculate the magnetic susceptibility
and the diamagnetic electric current distribution for graphene flakes
in Sec.\ \ref{sec_mag} and \ref{sec_cur}, respectively.
We make a quantitative comparison
between the orbital magnetism and the spin magnetism
in Sec.\ \ref{sec_spin}. We argue the magnetic-field alignment effect
in Sec.\ \ref{sec_align} and present
a brief conclusion in Sec.\ \ref{sec_conc}.


\section{Formulations}
\label{sec_form}

Graphene is composed of a honeycomb lattice of carbon
atoms, where a unit cell contains A and B sublattices.
We consider four different atomic configurations of graphene flakes
as shown in Fig.\ \ref{fig_flake},
which are characterized by hexagonal or trigonal shape
and by armchair or zigzag edge termination.
For each case, we range the system size from a few nm to a few tens of nm.
We describe the motion of graphene electrons 
using the nearest-neighbor tight-binding model for $p_z$ atomic orbitals.
The Hamiltonian is written as
\begin{align}
H=-\gamma_0\sum_{\langle n,m\rangle}e^{i2\pi \phi_{nm}}c_n^{\dagger}c_m, \label{hami}
\end{align}
where $-\gamma_0$ is the transfer integral, $c_n^\dagger$ is 
the creation operator at the site $n$, and
$\langle n,m\rangle$ represents summation over all 
nearest-neighbor sites.
The parameter $\gamma_0$ was experimentally estimated in the
bulk graphite as $\gamma_0 \approx 3$eV. 
\cite{Dresselhaus_and_Dresselhaus_2002a}
The system is under a uniform magnetic field $\bm{B}$ perpendicular to
the graphene plane, 
which is incorporated by the Peierls phase $\phi_{nm}$,
\begin{align}
\phi_{nm}=\frac{e}{ch}\int_n^m\mathrm{d}\bm{\ell}\cdot\bm{A},
\end{align}
where $\bm{A}(\bm{r})$ is the vector potential giving the magnetic field
by $\bm{B}=\nabla\times\bm{A}$.

For each single graphene flake, we diagonalize Hamiltonian Eq.\ (\ref{hami}) 
and obtain a set of eigenenergies $\e_i$.
The thermodynamical potential at temperature $T$ and 
chemical potential $\mu$ is written as
\begin{align}
\Omega=-k_{\mathrm{B}}T\sum_i\ln\left\{1+\exp[(\mu-\varepsilon_i)/k_{\mathrm{B}}T]\right\}. \label{omega}
\end{align}
The magnetic susceptibility per unit area is given by
\begin{align}
\chi=-\frac{1}{S}\left(\frac{\partial^2\Omega}{\partial B^2}\right)_{\mu,T}\biggr|_{B=0}, \label{chi}
\end{align}
where $S$ is the area of the system.
To calculate this, we obtain the eigenenergies at zero magnetic field 
and a small finite magnetic field,
and numerically calculate the derivative of the thermodynamic
potential.

The electric current from the site $m$ to $n$ is calculated by 
an operator,
\begin{align}
&J_{nm}=-i\frac{e\gamma_0}{\hbar}\left(e^{i2\pi\phi_{nm}}c_n^\dagger
 c_m-\mathrm{h.c.}\right).
\label{Jnm}
\end{align}
We obtain the expectation value of $J_{mn}$ for each bond
using the eigenstates at a sufficiently weak magnetic field,
where the current amplitude behaves linearly to $B$.


For the later references,
let us review the orbital magnetism of the bulk graphene. 
The low-energy physics of graphene electrons can be effectively 
described by the continuum massless Dirac Hamiltonian 
\cite{Wallace_1947a,McClure_1956a,Slonczewski_and_Weiss_1958a,DiVincenzo_and_Mele_1984a,Semenoff_1984a,Ando_2005a,Shon_and_Ando_1998a}
and the orbital susceptibility is calculated for this model as
\cite{McClure_1956a,Safran_and_DiSalvo_1979a,Koshino_and_Ando_2007b}
\begin{align}
\chi_\mathrm{eff}(\mu;T)&=
-g_v g_s\frac{e^2v^2}{24\pi c^2}\frac{1}{k_\mathrm{B}T\cosh^2[\mu/(2k_\mathrm{B}T)]}.
\label{bd}
\end{align}
where $g_v=g_s=2$ are the valley and spin degeneracies, respectively,
$v$ is the band velocity 
related to the transfer integral by $\hbar v=\sqrt{3}a\gamma_0/2$,
and $a \approx 0.246$nm is the lattice constant of graphene.
At $T=0$, Eq.\ (\ref{bd}) becomes a delta function,
\begin{align}
\chi_\mathrm{eff}(\mu;T=0)=-g_v g_s\frac{e^2v^2}{6\pi c^2}\delta(\mu).
\end{align}

\section{Magnetic susceptibility}
\label{sec_mag}

\begin{figure*}
\begin{center}
\leavevmode\includegraphics[width=0.9\hsize]{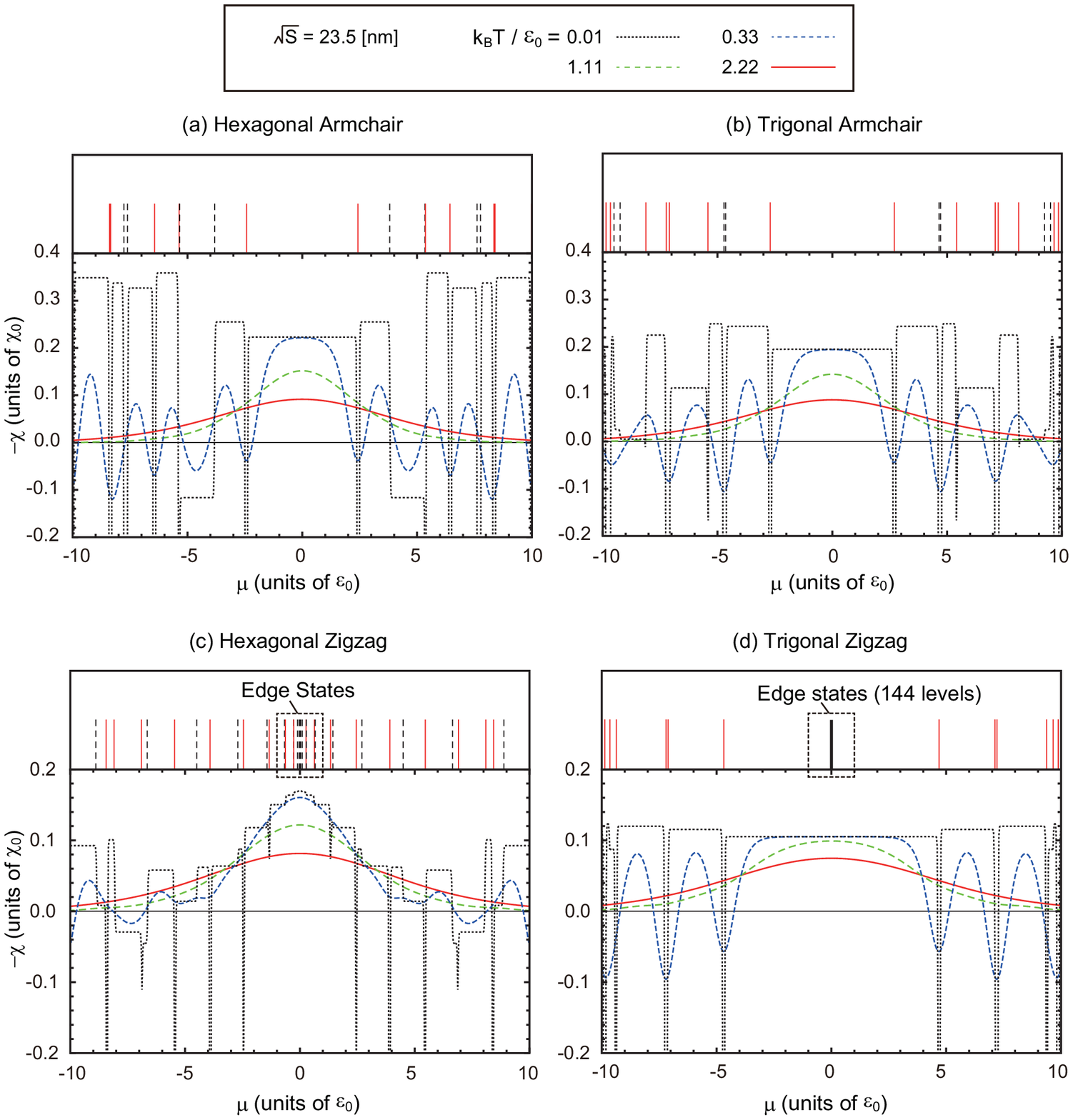}
\end{center}
\caption{Orbital magnetic susceptibility as a function
of $\mu$ in (a) hexagonal armchair, (b) trigonal armchair,
(c) hexagonal zigzag, and (d) trigonal zigzag graphene flakes
with the size of $\sqrt{S} \approx 23.5{\rm nm}$.
In each figure, the upper panel presents the energy spectrum,
where dashed (black) lines represent non-degenerate levels,
and solid (red) lines two-fold degenerate levels.
}
\label{mag_mu}
\end{figure*}

Fig.\ \ref{mag_mu} shows the susceptibility
against the chemical potential
for four types of the graphene flakes with several different 
temperatures.
The areas of the flakes are taken to be nearly equal to
$S\approx (23.5\mathrm{nm})^2$, which includes
$1.1\times 10^4$ of hexagonal unit cells.
The horizontal and vertical axes are scaled by 
\begin{eqnarray}
&& \e_0=\frac{\hbar v}{\sqrt{S}}, \\
&& \chi_0=\frac{g_v g_s e^2v^2}{6\pi c^2 \e_0},
\end{eqnarray}
respectively.
$\e_0$ represents the energy scale in the Dirac cone
associated with the length scale $\sqrt{S}$.
We also calculated the susceptibility for different system sizes 
and found that for each of four types, the susceptibility
and the level structure plotted in this scale becomes almost universal 
as long as $\sqrt{S} \gg a$.
This is naturally expected from the fact that
the low-energy physics are well described by the
Dirac Hamiltonian. 

Upper figure in each panel
presents the energy level structure at $B=0$,
where dashed (black) lines are non-degenerate levels,
and solid (red) lines are two-fold degenerate levels.
In the low temperature regime, $k_{\mathrm{B}}T \ll \e_0$,
we observe that
the susceptibility abruptly changes at every single energy level,
and in particular, it exhibits sharp spikes toward the paramagnetic
direction (downward in the figure) at two-fold degenerate levels.
This is because the degenerate states, 
having opposite magnetic moments, 
split linearly in magnetic field just like spin Zeeman splitting,
and induce paramagnetism in an analogous way 
to spin Pauli paramagnetism.
The contribution to the orbital susceptibility (per area)
from the degenerate states at $E_0$ is written as
\begin{eqnarray}
 \chi = \frac{2 m^2}{S}\delta (\mu - E_0),
\label{eq_chi_deg}
\end{eqnarray}
where $\pm m$ is the magnetic moments of the doublet.
The typical magnitude of $m$ is shown to be
$\sqrt{S} ev/c$, which is 
the only magnetic-moment scale in the massless Dirac system.


The major difference 
between armchair flakes and zigzag flakes
comes from the existence of 
the zero-energy edge states peculiar to the zigzag edge.
\cite{Fujita_et_al_1996a,Nakada_et_al_1996a,Wakabayashi_et_al_1999a}
In the triangular zigzag flake, Fig.\ \ref{mag_mu}(d),
there are a number of energy levels exactly at zero energy,\cite{Ezawa_2007a}
of which wavefunctions are shown to be localized
at the edge, and the degeneracy is the order of $\sim \sqrt{S}/a$. 
Remarkably, the susceptibility in the low temperature regime
is completely flat at these levels, meaning that the edge states have 
absolutely no contribution to the orbital magnetism.
This is simply because the edge states are locked 
to zero energy even in the presence of magnetic field,
and never participate in the total energy change.
In the hexagonal zigzag flake, Fig.\ \ref{mag_mu}(c),
on the other hand, the edge levels slightly shift from zero energy,
leading to some small contributions to the orbital susceptibility.
The energy shift arises because
the edge states on neighboring sides of the hexagon
always reside at different sublattices, and they are 
hybridized by some finite matrix element including $\gamma_0$.
Nevertheless, the edgestates do not play a significant role
in the overall behavior of the orbital magnetism.

\begin{figure}
\begin{center}
\leavevmode\includegraphics[width=0.95\hsize]{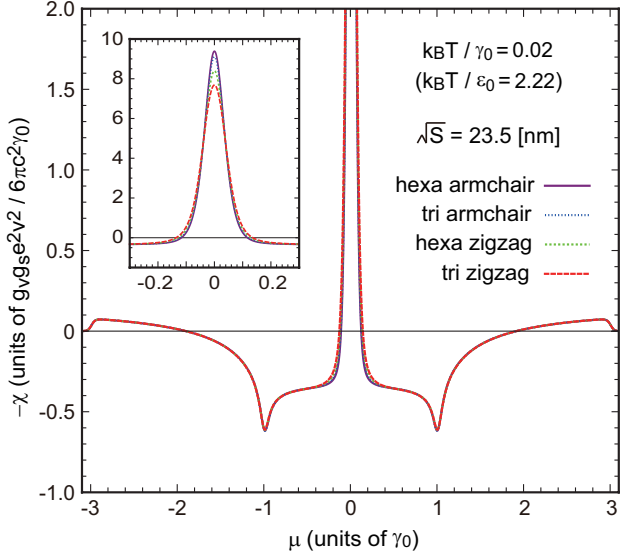}
\end{center}
\caption{Extended plot 
of the susceptibility curves in Fig.\ \ref{mag_mu}
over the whole band region, for the four types of graphene flakes
at $k_\mathrm{B}T/\e_0= 2.22$.
The energy axis is now scaled by absolute unit $\gamma_0$,
in which the temperature amounts to $k_\mathrm{B}T/\gamma_0=0.02$.
Inset shows the detail of the central peak.
}
\label{ha_mag_mu}
\end{figure}

\begin{figure}
\begin{center}
\leavevmode\includegraphics[width=0.9\hsize]{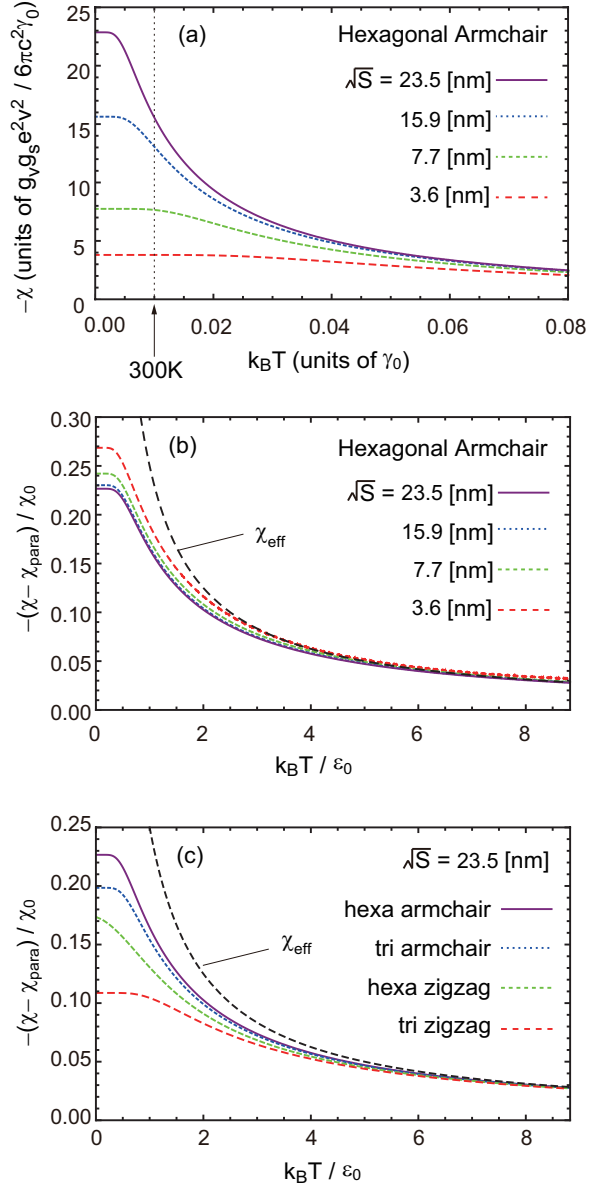}
\end{center}
\caption{(a) $\chi(\mu=0;T)$
of hexagonal armchair flakes with several different sizes,
plotted in the absolute units $\gamma_0$ 
and $g_v g_s e^2v^2/(6\pi c^2 \gamma_0)$.
(b) $\chi(\mu=0;T) - \chi_{\rm para}$ of the same systems,
plotted in the relative units $\e_0$ 
and $\chi_0$.
(c)  Plot similar to (b) for
different types of graphene flakes
with the size $\sqrt{S} \approx 23.5{\rm nm}$.
}
\label{mag_t}
\end{figure}


As the temperature increases, the spikes and steps 
in the susceptibility are
smeared out into an oscillatory curve.
The oscillation eventually disappears 
in $k_B T /\e_0 > \sim 1$,
leaving a single diamagnetic peak at the Dirac point,
which corresponds to the thermally-broadened delta-function in the
bulk limit, Eq.\ (\ref{bd}).
In Fig.\ \ref{ha_mag_mu}, we present an extended plot 
of the susceptibility curves at $k_\mathrm{B}T/\e_0= 2.22$ 
over the whole band region, for different types of graphene flakes
with $\sqrt{S}\approx 23.5\mathrm{nm}$.
The energy axis is now scaled by absolute unit $\gamma_0$.
We see that 
the finite-size effect almost vanishes in this temperature regime,
giving a universal curve independent of the atomic configuration.
The curves still slightly differ near the central peak,
because the level spacing around the Dirac point, which is about
$\sim \e_0$, is not completely negligible compared to $k_\mathrm{B} T$
at this particular system size.
This small variance would vanish in larger flakes which satisfy
$k_\mathrm{B} T \gg \e_0$.

The curve 
is characterized by the strong diamagnetic peak at the Dirac point
and some smaller structures off the Dirac point.
The contribution from the lower-half spectrum 
adds up to a paramagnetic offset
to the central diamagnetic peak. Namely,
the susceptibility near $\mu = 0$ is approximately written as
\begin{align}
\chi(\mu;T) \approx \chi_\mathrm{eff}(\mu;T)+\chi_\mathrm{para},
\end{align}
where $\chi_\mathrm{eff}$ is given by Eq.\ (\ref{bd}), and
\begin{eqnarray}
\chi_\mathrm{para}\approx 0.37 \times 
\frac{g_v g_s e^2v^2}{6\pi c^2\gamma_0}. 
\end{eqnarray}
The offset $\chi_\mathrm{para}$ is much smaller
than the height of the central peak
$\approx g_v g_s e^2v^2/(24\pi c^2k_\mathrm{B}T)$,
since $k_\mathrm{B}T$ is usually much smaller than $\gamma_0$.
Outside the Dirac cone region, we see tiny Landau diamagnetism
in the quadratic band bottom at $\mu = \pm 3 \gamma_0$,
and paramagnetism 
around the van Hove singularity at $\mu = \pm \gamma_0$.
\cite{Safran_and_DiSalvo_1979a,Blinowski_and_Rigaux_1984a,Saito_and_Kamimura_1986a}


To analyze the size dependence quantitatively,
we plot $\chi(\mu=0;T)$ 
of hexagonal armchair flakes with different sizes
in Fig.\ \ref{mag_t}(a) and (b).
The panels (a) and (b) present the same information
but in different fashions:
(a) shows $\chi$ in the absolute units $\gamma_0$ 
and $g_v g_s e^2v^2/(6\pi c^2 \gamma_0)$
for horizontal and vertical axes, respectively,
while (b) shows $\chi - \chi_{\rm para}$
i.e., the contribution from the Dirac cone,
with relative units $\e_0$ and $\chi_0$ depending on 
the system size.
In (b), we see that the curve 
converges to a single universal curve as the size increases,
indicating that the physics there is well described by 
effective Dirac equation.
The susceptibility approaches the bulk limit $\chi_{\rm eff}$
in the high temperature region $k_\mathrm{B}T \gg \e_0$, whereas
in $k_\mathrm{B}T <\sim \e_0$
it deviates from $\chi_{\rm eff}$ and reaches
some finite maximum value.
When we consider the susceptibility of a single graphene flake,
$\chi S$, at a fixed absolute temperature,
it scales in proportion to $\chi_0 S \propto S^{3/2}$ 
in the low-temperature regime $k_\mathrm{B} T <\sim \e_0$,
while it is just proportional to $S$ 
in the high-temperature regime $k_\mathrm{B}T \gg \e_0$
where $\chi$ is equal to the bulk limit.

The detail of the universal curve in  Fig.\ \ref{mag_t}(b) depends 
on the flake shape and the edge configuration.
In Fig.\ \ref{mag_t}(c), 
we present plots similar to Fig.\ \ref{mag_t}(b)
for four different types of graphene flakes
with the same size $\sqrt{S}\approx 23.5\mathrm{nm}$,
which is sufficiently large to achieve the universal limit.
In low temperatures, the susceptibility tends to be larger in 
an armchair flake than in a zigzag flake,
and larger in a hexagonal flake than a trigonal flake.
In the high temperature region, on the other hand,
all the curves approaches the same bulk limit.
A similar edge dependence was previously found in graphene ribbons,
where armchair ribbons generally exhibit
larger diamagnetism than zigzag ribbons.
\cite{Wakabayashi_et_al_1999a,Ominato_and_Koshino_2012a}

\section{Diamagnetic current distribution}
\label{sec_cur}

\begin{figure*}
\begin{center}
\leavevmode\includegraphics[width=0.9\hsize]{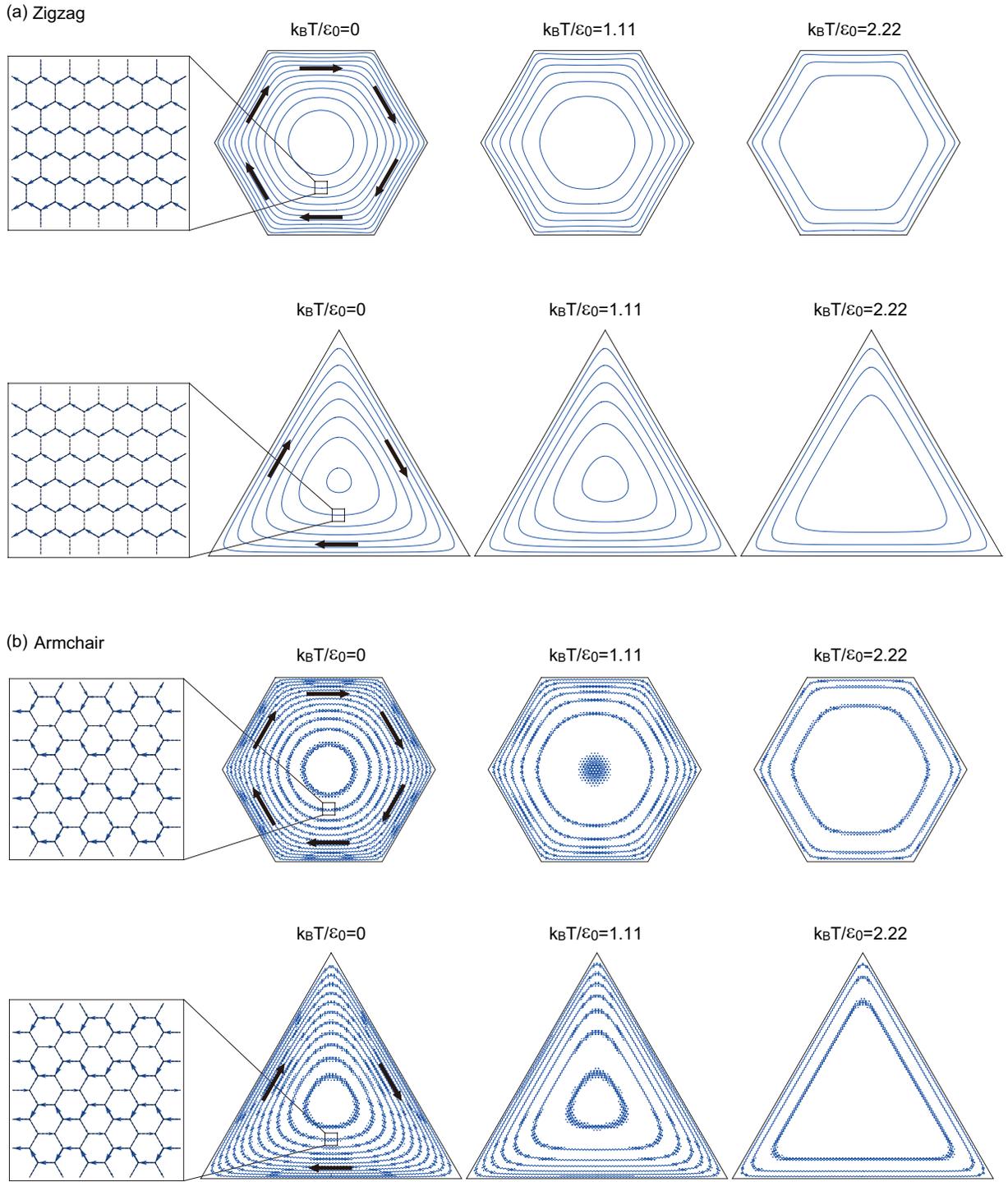}
\end{center}
\caption{
Diamagnetic current distribution 
in different types of graphene flakes
of the same size $\sqrt{S}\approx 23.5\mathrm{nm}$
at several different temperatures.
Continuous flux lines are
obtained by smoothing the original discrete 
current on each bond, which is shown in the left.
}
\label{flux}
\end{figure*}

\begin{figure}
\begin{center}
\leavevmode\includegraphics[width=0.9\hsize]{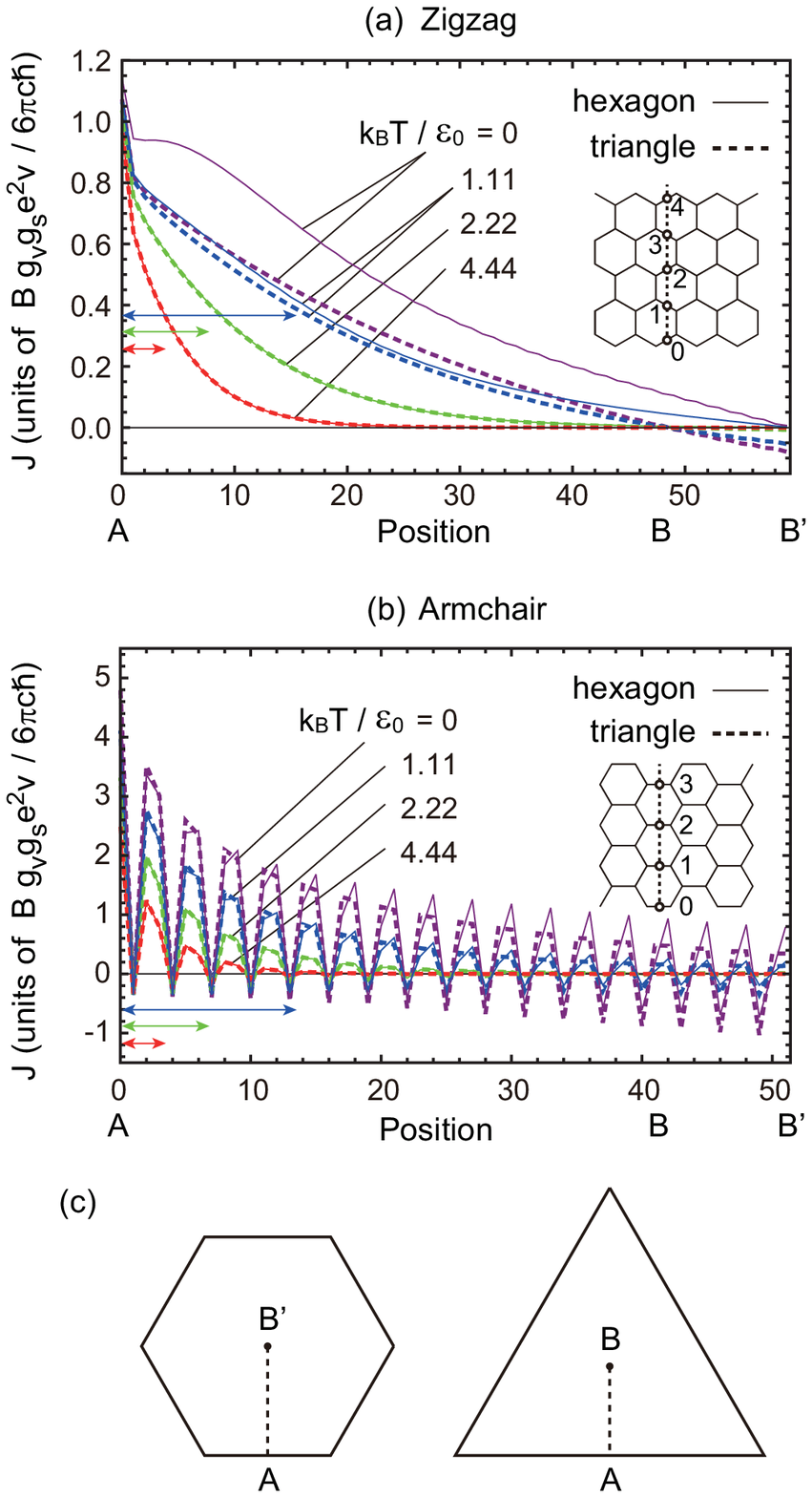}
\end{center}
\caption{
Electric current as a function of position from the boundary to the center, 
in (a) the zigzag and (b) armchair flake
with $\sqrt{S}\approx 23.5\mathrm{nm}$..
The position is labeled by the bond index defined in the inset,
and A and B (B') represent the border and the center of
triangle (hexagon), respectively, which are specified in (c).
Horizontal arrows indicate $\lambda_{\rm edge}$
for $k_\mathrm{B}T/\e_0 = 1.11,2.22$ and $4.44$.
}
\label{cur}
\end{figure}

Fig.\ \ref{flux} shows the diamagnetic current distribution induced
by the magnetic field in the four types of graphene flakes
of the size $\sqrt{S}\approx 23.5\mathrm{nm}$
at several different temperatures.
To visualize the global current circulation,
we illustrate continuous flux lines 
obtained by smoothing the original discrete 
current $J_{mn}$ on each bond, which is shown in the left inset.
Specifically, we find a certain potential function $\Psi$
which satisfies $\bm{J} = \bm{e}_z \times \nabla \Psi$,
and obtain the equi-potential lines of $\Psi$
as the current flux lines.
At zero temperature, the flux circulates entirely on the system
reflecting the absence of
the characteristic wave length in graphene.
As temperature becomes higher, it is going to be localized near the edge.
The current circulation of zigzag and armchair graphene flakes
are globally similar, but the flux lines of armchair flakes
exhibit some roughness while it is not observed in zigzag flakes.
This actually 
corresponds to the atomic-scale
current circulation in the Kekul\'{e} pattern
seen in the original current map (left inset),
\cite{Wakabayashi_et_al_1999a}
which is caused by the inter-valley (between $K$ and $K'$) 
hybridization peculiar to the armchair edge.

Fig.\ \ref{cur} shows the detailed plots of the electric current 
as a function of position from the boundary to the center, 
calculated for (a) the zigzag and (b) armchair flakes.
The position is labeled by the bond index defined in the inset,
and A and B (B') correspond to the edge and the center of
triangle (hexagon), respectively, which are depicted in Fig.\ \ref{cur}(c).
The current distribution is more localized to the edge  
when $T$ becomes higher,
and the typical depth of the edge current is characterized by 
\begin{align}
\lambda_\mathrm{edge}=\frac{\hbar v}{2\pi k_\mathrm{B}T},
\end{align}
 in accordance with the result for graphene ribbons.
\cite{Ominato_and_Koshino_2012a}

The current distribution in the atomic scale strongly depends
on the edge type. We can show that, however,
the integrated edge current approximates $c \chi_{\rm eff} B$ 
independently of the edge type,
in the high temperature regime $k_\mathrm{B}T/\e_0 \gtrsim 2$.
This is consistent with the fact that
the orbital susceptibility is then given by the bulk limit
regardless of the atomic configuration.
When comparing hexagonal and triangular flakes of the same edge type, 
we see that the curves are almost completely equivalent 
in $k_\mathrm{B}T/\e_0 \gtrsim 2$.
This suggests that the edge current distribution in high temperature
is solely determined by the local edge configuration,
independently of the global shape.

\section{Comparison to spin paramagnetism}
\label{sec_spin}


The orbital magnetism always competes with the 
spin paramagnetism which has been neglected so far.
When we include spin Zeeman splitting, 
each  spin-less  energy level at $E_0$ acquires
the Pauli paramagnetism
\begin{eqnarray}
 \chi_{\rm Pauli} = \frac{1}{S} \left(\frac{g}{2}\mu_\mathrm{B}\right)^2 2 \delta (\mu - E_0),
\end{eqnarray}
where $g \sim 2 $ is the g factor for a graphene electron,
$\mu_\mathrm{B} = e\hbar/(2m_0 c)$ is the Bohr magneton
and $m_0$ is the bare electron mass.
This is similar to the orbital contribution of Eq.\ (\ref{eq_chi_deg})
for doubly degenerate levels,
while the orbital magnetic moment $m$ there
is now replaced with $g\mu_\mathrm{B}/2$.
In the flakes of $S>(1{\rm nm})^2$,
$\mu_\mathrm{B}$ is much smaller than the typical magnitude of $m$, 
which is $\approx \sqrt{S} ev/c$, 
suggesting that 
the Pauli paramagnetic effect is 
typically much smaller than the orbital effect.
This in contrast to conventional electron systems 
where orbital magnetic moment 
and spin magnetic moment are both of the order of $\mu_\mathrm{B}$.
\cite{Kubo_and_Obata_1956a}

In a zigzag graphene flake,
the highly-degenerate edge states at zero energy
give exceptionally large Pauli paramagnetism.
The contribution is written as
\begin{eqnarray}
 \chi_{\rm Pauli} = 
\frac{N_{\rm edge}}{S}
 \left(\frac{g}{2}\mu_\mathrm{B}\right)^2
2 \delta (\mu),
\end{eqnarray}
where $N_\mathrm{edge}(\sim\sqrt{S}/a)$ is the number of edge states
per spin.
In the low-temperature regime such that $k_\mathrm{B} T \ll \e_0$,
this is dominant over the orbital effect near zero energy,
since the orbital susceptibility does not diverge 
at edge states as already shown.
In high-temperature regime $k_\mathrm{B} T \gg \e_0$,
the delta-function is thermally broadened 
and it should be compared to the bulk orbital susceptibility
$\chi_\mathrm{eff}$, Eq.\ (\ref{bd}).
The ratio between two opposite components approximates
\cite{Ominato_and_Koshino_2012a}
\begin{align}
\bigg|\frac{\chi_\mathrm{Pauli}}{\chi_\mathrm{eff}}\bigg|
\sim \frac{3\pi}{g_vg_s}
\left(\frac{g}{2}
\frac{\hbar}{m_0 va}\right)^2 \frac{a}{\sqrt{S}}
\sim 0.4\times\frac{a}{\sqrt{S}},
\end{align}
so that the Pauli paramagnetism 
is negligible in a large flake with $\sqrt{S}\gg a$.

It should be noted that graphene flakes 
may have lattice vacancies and/or adatoms depending on the 
experimental condition, and the impurity levels given by these defects
contribute to additional Pauli paramagnetism.
Moreover, we remark that several experimental studies reported
the evidence of ferromagnetic spin ordering in 
graphene-based materials. 
\cite{Esquinazi_et_al_2002a,Esquinazi_et_al_2003a,Enoki_and_Takai_2009a,Wang_et_al_2009a}
The origin of the spontaneous magnetism is still under debate,
while it is supposed to be caused by
the atomic defects, grain boundaries, 
and highly-degenerate edge states.
\cite{Wakabayashi_et_al_1999a,Fernandez_and_Palacios_2007a,Wang_and_Meng_and_Kaxiras_2008a}



\section{Magnetic field alignment of graphene flakes}
\label{sec_align}

\begin{figure}
\begin{center}
\leavevmode\includegraphics[width=1.0\hsize]{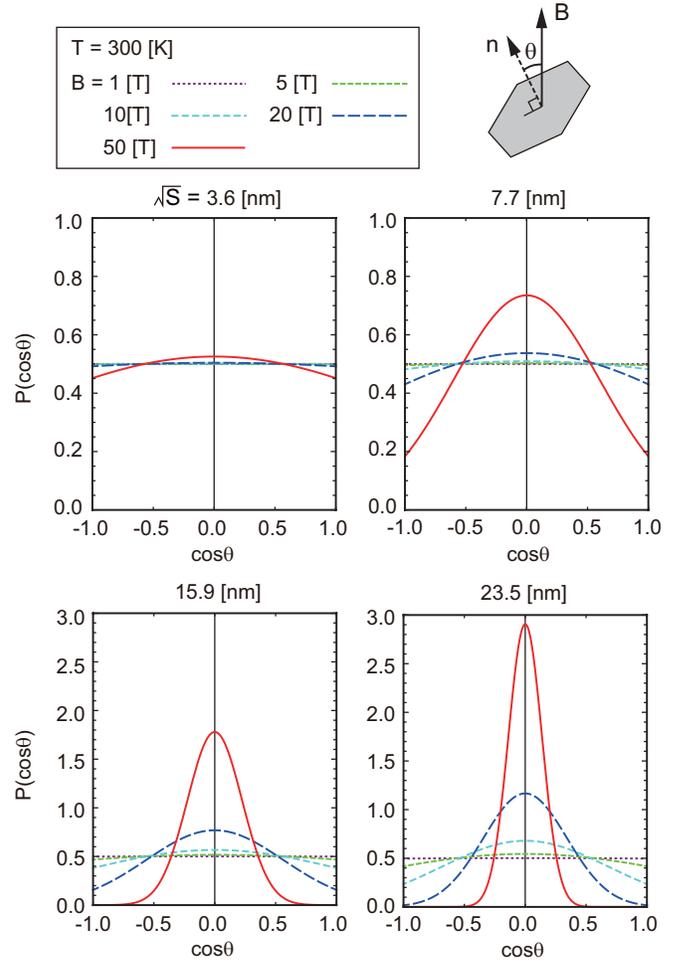}
\end{center}
\caption{
Angle distribution of hexagonal armchair flakes 
in magnetic fields at $T=300$K.
}
\label{distribution}
\end{figure}

The diamagnetism of graphene can be possibly observed 
using the magnetic-field alignment of 
graphene nanoflakes dissolved in a solvent,
similarly to the experiments for the carbon nanotube.\cite{Zaric_et_al_2004a}
In a magnetic field,
the graphene flakes tends to be oriented parallel to the field direction,
because the field component 
penetrating the graphene plane raises the total energy due to the diamagnetism.
If we assume that the graphene flakes are planer and rigid,
the condition to achieve the alignment is roughly estimated as
\begin{align}
\frac{1}{2}\chi B^2S\gtrsim k_\mathrm{B}T. \label{align}
\end{align}
For the graphene flakes $\sqrt{S}\approx 23.5\mathrm{nm}$  at $T=300$K, 
for example,
the required field becomes $B \gtrsim 9$T.

We calculate the angle distribution of graphene flakes with 
various sizes using the Maxwell-Boltzmann statistics.
In the thermal equilibrium, 
the probability that the normal of the graphene plane 
is inclined from the magnetic field by $\theta$ to $\theta + d\theta$ 
is written as $P(\cos\theta) d(\cos\theta)$,
where
\begin{align}
P(\cos\theta)=\frac{\exp[-\beta U(\cos\theta)]}
{\int^1_{-1}\exp[-\beta U(\cos\theta)]\mathrm{d}(\cos\theta)},
\end{align}
with $U(\cos\theta)=-(1/2)\chi S B^2\cos^2\theta$
and $\beta = 1/(k_\mathrm{B}T)$. 
Fig.\ \ref{distribution} plots the distribution function
$P(\cos\theta)$ 
calculated for hexagonal armchair flakes with
several sizes at $T=300$K,  
using $\chi$ in Fig.\ \ref{mag_t}(a).
We see that the alignment occurs more strongly in larger flakes,
because the magnetization of a single flake, $\chi S B$,
is greater for larger $S$. 
Note that it is not only due to a linear factor $S$,
but also because $\chi$ increases in larger $S$ 
as shown in Fig.\ \ref{mag_t}(a).

\section{CONCLUSION}
\label{sec_conc}

We have studied the orbital diamagnetism
of the graphene flakes with various shapes and edge configurations
using the tight-binding approximation.
We found that the behavior is significantly different 
depending on the relative magnitude of 
the thermal broadening energy $k_\mathrm{B}T$
to the typical energy level spacing $\e_0 = \hbar v / \sqrt{S}$.
In the low-temperature regime where $k_\mathrm{B}T \ll \e_0$,
the susceptibility as a function of Fermi energy
rapidly changes between diamagnetism
and paramagnetism in accordance with the level structure
depending on the specific atomic structure of the flake.
The susceptibility at the zero Fermi energy 
is found to be generally larger in armchair flakes than in zigzag flakes,
and larger in hexagonal flakes than trigonal flakes.
In the high-temperature regime $k_\mathrm{B}T \gtrsim 2\e_0$, on the other hand,
the discrete structures due to the finite-size effect are all gone,
and the susceptibility approximates the bulk limit
independently of the shape and the edge configuration of the flake.
Considering $\e_0$ is written as
$8000{\rm [K]}/\sqrt{S}{\rm [nm]}$ 
using the graphene's band velocity,
we find that the room temperature belongs to the low temperature regime
for a flake of a few nanometer, 
while it is in the high temperature regime
for a flake more than 50nm.

In the low-temperature regime,
the diamagnetic current circulates entirely
on the graphene flakes, 
reflecting the absence of characteristic length scale.
As the temperature increases, the current gradually becomes to
circulate only near the edge, with the characteristic depth of
$\lambda_\mathrm{edge}=\hbar v/2\pi k_\mathrm{B}T$.
The local current distribution along the cross section perpendicular to the
boundary is insensitive to the global shape of the flake, 
but significantly different between armchair and zigzag edges.

We predict that the diamagnetism of graphene can be possibly observed 
using the magnetic-field alignment of graphene flakes.
We estimated the angle distribution at various magnetic fields, 
and found that a strong alignment
can be realized in the feasible magnetic field range
for flakes of $S \gtrsim (10{\rm nm})^2$.

\section*{ACKNOWLEDGMENTS}

Authors thank helpful discussions with Toshiaki Enoki and
Kazuyuki Takai.
This work was supported by Grants-in-Aid for Scientific
Research No.24740193 from JSPS.

\end{document}